\documentclass[twoside,twocolumn,9pt]{article}
\usepackage{extsizes}
\usepackage[super,sort&compress,comma]{natbib} 
\usepackage[version=3]{mhchem}
\usepackage[left=1.5cm, right=1.5cm, top=1.785cm, bottom=2.0cm]{geometry}
\usepackage{balance}
\usepackage{times,mathptmx}
\usepackage{sectsty}
\usepackage{graphicx} 
\usepackage{lastpage}
\usepackage[format=plain,justification=justified,singlelinecheck=false,font={stretch=1.125,small,sf},labelfont=bf,labelsep=space]{caption}
\usepackage{float}
\usepackage{fancyhdr}
\usepackage{fnpos}
\usepackage[english]{babel}
\addto{\captionsenglish}{%
  
}
\usepackage{array}
\usepackage{droidsans}
\usepackage{charter}
\usepackage[T1]{fontenc}
\usepackage[usenames,dvipsnames]{xcolor}
\usepackage{setspace}
\usepackage[compact]{titlesec}
\usepackage{hyperref}

\usepackage{epstopdf}

\definecolor{cream}{RGB}{222,217,201}

\begin{document}

\pagestyle{fancy}
\thispagestyle{plain}
\fancypagestyle{plain}{

\fancyhead[C]{\includegraphics[width=18.5cm]{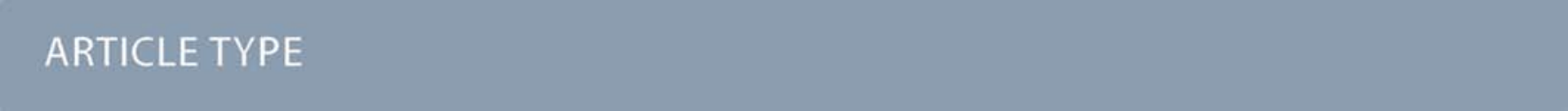}}
\fancyhead[L]{\hspace{0cm}\vspace{1.5cm}\includegraphics[height=30pt]{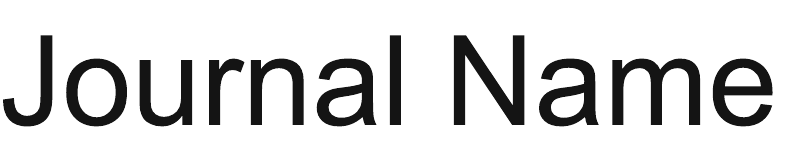}}
\fancyhead[R]{\hspace{0cm}\vspace{1.7cm}\includegraphics[height=55pt]{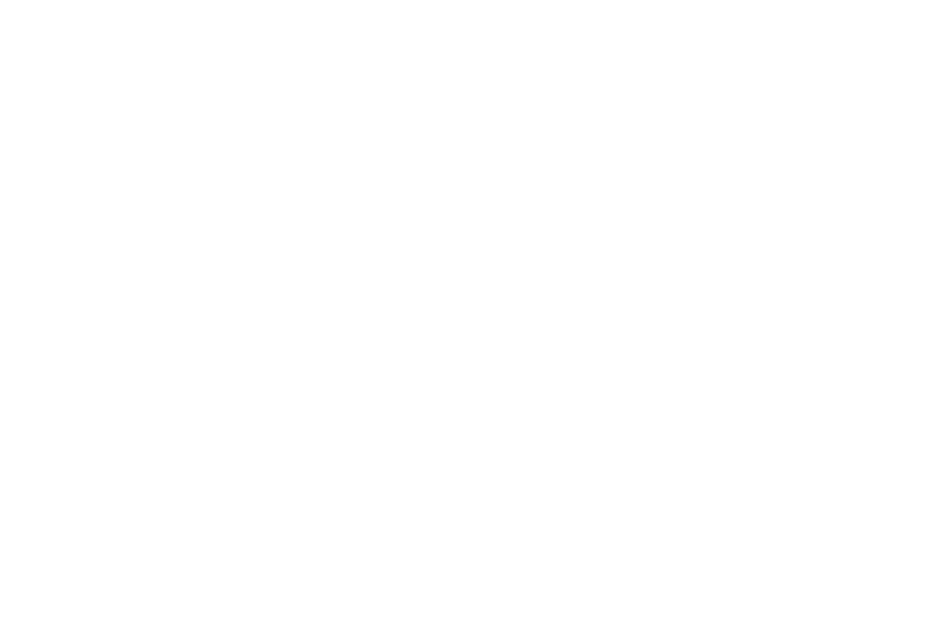}}
\renewcommand{\headrulewidth}{0pt}
}

\makeFNbottom
\makeatletter
\renewcommand\LARGE{\@setfontsize\LARGE{15pt}{17}}
\renewcommand\Large{\@setfontsize\Large{12pt}{14}}
\renewcommand\large{\@setfontsize\large{10pt}{12}}
\renewcommand\footnotesize{\@setfontsize\footnotesize{7pt}{10}}
\makeatother

\renewcommand{\thefootnote}{\fnsymbol{footnote}}
\renewcommand\footnoterule{\vspace*{1pt}%
\color{cream}\hrule width 3.5in height 0.4pt \color{black}\vspace*{5pt}} 
\setcounter{secnumdepth}{5}

\makeatletter 
\renewcommand\@biblabel[1]{#1}            
\renewcommand\@makefntext[1]%
{\noindent\makebox[0pt][r]{\@thefnmark\,}#1}
\makeatother 
\renewcommand{\figurename}{\small{Fig.}~}
\sectionfont{\sffamily\Large}
\subsectionfont{\normalsize}
\subsubsectionfont{\bf}
\setstretch{1.125} 
\setlength{\skip\footins}{0.8cm}
\setlength{\footnotesep}{0.25cm}
\setlength{\jot}{10pt}
\titlespacing*{\section}{0pt}{4pt}{4pt}
\titlespacing*{\subsection}{0pt}{15pt}{1pt}

\fancyfoot{}
\fancyfoot[LO,RE]{\vspace{-7.1pt}\includegraphics[height=9pt]{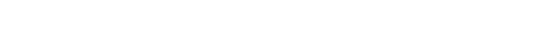}}
\fancyfoot[CO]{\vspace{-7.1pt}\hspace{13.2cm}\includegraphics{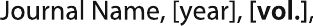}}
\fancyfoot[CE]{\vspace{-7.2pt}\hspace{-14.2cm}\includegraphics{rf}}
\fancyfoot[RO]{\footnotesize{\sffamily{1--\pageref{LastPage} ~\textbar  \hspace{2pt}\thepage}}}
\fancyfoot[LE]{\footnotesize{\sffamily{\thepage~\textbar\hspace{3.45cm} 1--\pageref{LastPage}}}}
\fancyhead{}
\renewcommand{\headrulewidth}{0pt} 
\renewcommand{\footrulewidth}{0pt}
\setlength{\arrayrulewidth}{1pt}
\setlength{\columnsep}{6.5mm}
\setlength\bibsep{1pt}

\makeatletter 
\newlength{\figrulesep} 
\setlength{\figrulesep}{0.5\textfloatsep} 

\newcommand{\topfigrule}{\vspace*{-1pt}%
\noindent{\color{cream}\rule[-\figrulesep]{\columnwidth}{1.5pt}} }

\newcommand{\botfigrule}{\vspace*{-2pt}%
\noindent{\color{cream}\rule[\figrulesep]{\columnwidth}{1.5pt}} }

\newcommand{\dblfigrule}{\vspace*{-1pt}%
\noindent{\color{cream}\rule[-\figrulesep]{\textwidth}{1.5pt}} }

\makeatother

\twocolumn[
  \begin{@twocolumnfalse}
\vspace{3cm}
\sffamily
\begin{tabular}{m{4.5cm} p{13.5cm} }

\includegraphics{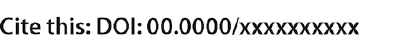} & \noindent\LARGE{\textbf{A simple neural network implementation of generalized solvation free energy for assessment of protein structural models $^\dag$}} \\
\vspace{0.3cm} & \vspace{0.3cm} \\

 & \noindent\large{Shiyang Long,$^{\ast}$\textit{$^{a}$} and Pu Tian\textit{$^{b}$}} \\

\includegraphics{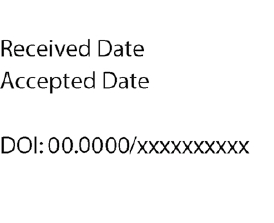} & \noindent\normalsize{Rapid and accurate assessment of protein structural models is essential for protein structure prediction and design. Great progress has been made in this regard, especially by recent development of ``knowledge-based'' potentials. Various machine learning based protein structural model quality assessment was also quite successful. However, performance of traditional ``physics-based'' potentials have not been as effective. Based on analysis of computational limitations of present solvation free energy formulation, which partially underlies unsatisfactory performance of ``physics-based'' potentials, we proposed a generalized sovation free energy (GSFE) framework. GSFE is intrinsically flexible for multi-scale treatments and is amenable for machine learning implementation. In this framework, each physical comprising unit of a complex molecular system has its own specific solvent environment. One distinctive feature of GSFE is that high order correlations within selected solvent environment might be captured through machine learning, in contrast to present empirical potentials (both ``knowledge-based'' and ``physics-based'') that are mainly based on pairwise interactions. Finally, we implemented a simple example of backbone and side-chain orientation based residue level protein GSFE with neural network, which was found to have competitive performance when compared with highly complex latest ``knowledge-based'' atomic potentials in distinguishing native structures from decoys.} \\

\end{tabular}

 \end{@twocolumnfalse} \vspace{0.6cm}

  ]

\renewcommand*\rmdefault{bch}\normalfont\upshape
\rmfamily
\section*{}
\vspace{-1cm}


\footnotetext{\textit{$^{a}$~School of Chemistry, Jilin University}}
\footnotetext{\textit{$^{b}$~School of Life Science and School of Artificial Intelligence, Jilin University, 2699 Qianjin Street, Changchun, China 130012}}

\footnotetext{\dag~Electronic Supplementary Information (ESI) available: [the derivation steps of equation 8]. See DOI: 00.0000/00000000.}



\section{INTRODUCTION}
With time-consuming, expensive and sometimes extremely challenging high resolution experimental analysis, reliable and accurate computational protein structure prediction has been highly desired due to tremendous amount of protein sequences generated by present sequencing technologies\cite{Land2015}. The long lasting and steadily increasing interest in this regard was evidenced by the impressive development of the CASP (Critical Assessment of techniques for protein Structure Prediction, www.predictioncenter.org) community. The fundamental underlying assumption of protein folding was that native structures have the lowest free energy among all possible structural arrangements of residues/atoms in space\cite{Anfinsen223}. Unfortunately, quantum mechanically (or even atomically) rigorous calculation of free energy for any realistic protein molecule was intractable on the one hand, and unnecessary on the other hand. Development of approximate free energy estimators, widely addressed as potentials or scoring functions, was therefore essential for efficient protein structure prediction. Two basic steps of protein structure predictions were proposal/sampling and assessment of structural models, the later step was performed by comparing scores given by various methodologies, which could be classified into two major categories, ``knowledge-based'' (KB) and ``physics-based'' (PB)\cite{Skolnick2006} ones. It was widely realized that KB potentials were much more effective than PB potentials in recognizing native structures from decoys\cite{Pei2019}. 

A number of KB potentials were based on atom pair distances, possibly with various forms of orientation consideration\cite{Zhang2010,Chae2015,Hoque2016,Xu2017,Yu2019}. Reference state definition was acknowledged to be a major source of differences among them\cite{Zhang2010,Liu2014}. Additionally, definition of reference state was found to impact selection of cutoff distances\cite{Yao2017}. OPUS-CSF utilized some designed feature extracted from backbone segments of various lengths (5 to 11 residues) to describe extent of nativeness\cite{Xu2018} and it was further developed to include side chains\cite{Xu2019}. Utilization of rotameric states besides distances\cite{Park2014}, and voronoi contact suface\cite{Olechnovic2017} were also found to be quite successful. Explicit consideration of entropy in some models was found to be helpful\cite{Sankar2017, Wang2018}. These carefully handcrafted potentials have progressed steadily in performance and were relatively easy to explain physically. The limitation was that upgrading involved significant human intervention.

Many machine learning protein structural model quality assessment protocols have been developed with great success. Both single quality assessment models\cite{Cao2016a,Cao2016,Cao2017,Manavalan2017,Gao2018,Derevyanko2018,Pei2019} and meta-models\cite{Cao2015,Jing2017,Mulnaes2018} have been utilized. Multiple object optimization approach were also investigated\cite{Song2018,Song2018a}. No complex derivation of specific functional formulations were necessary. However, physical explanation was less straight forward for these machine learning based protocols.  

In this paper, we analyzed the computational limitations of present solvation free energy theoretical formulations underlying PB free energy estimation, and proposed an intrinsically multi-scale generalized solvation free energy (GSFE) framework to facilitate application of powerful machine learning optimization algorithms. Furthermore, we provided a simple neural network implementation of GSFE framework at residue level for assessment of backbone protein structural models (with side-chain orientation). Despite simplicity, this implementation was found to be competitive when compared with state-of-the-art KB methods that were significantly more complex. Further development of GSFE for quality assessment of protein structural models and protein design will be carried out in the near future. We emphasized here that traditional solvation free energy formulations remain powerful tools in tackling many problems. Our formulation of GSFE was one possible alternative way to facilitate utilization of machine learning optimization capacity while maintaining some level of physical explanation.

\section{METHODOLOGY}

\subsection{Computational limitation of present solvation free energy formulation}
It has been widely accepted that hydrophobic interactions are among the most important driving forces for protein folding\cite{Dill2012,Bellissent-Funel2016}. Therefore, solvation free energy has been an important topic with a long history of theoretical development\cite{Matubayasi}. Protein solvation free energy was itemized as an independent term as indicated by the following equation\cite{Matubayasi}:
\begin{align}
-k_BTlogP(S) = E_{Intra}(S) + \Delta\mu(S) + Constant
\end{align}
with $S$ being the configuration of protein atoms and $\Delta\mu$ being the solvation free energy when the protein is in configuration $S$, $P(S)$ being the probability that protein is in $S$ configuration. The constant term was determined by selection of reference state. This framework was conceptually helpful for understanding and also amenable for theoretical derivation. Most importantly, it was very convenient for us to explicitly treat many interesting and useful interfacial properties. These advantages explained its popularity. The solvation free enengy term $\Delta\mu$ was usually further divided into polar and nonpolar term. With implicit treatment of solvents (which was usually the case for protein structure prediction), calculation of polar term mainly involved solving Poisson-Boltzmann equation or its linearized approximations\cite{Wang2008}, but calculation of the nonpolar term was much more challenging\cite{Chen2008}. Additionally, direct and reliable separation of free energy of a protein molecular system into internal and solvation free energy experimentally was not an easy task for biomolecular systems. Therefore, in development of computational models according to this classical formulation, direct experimental validation for either term (intramolecular packing and solvation) was not readily available. Meanwhile, due to complexity of configurational sampling and electrostatic calculations, theoretical approximations were unavoidable and experimental validation therefore was essential for development of reliable computational methodologies. Present strategies of solving this dilemma has been to validate computational algorithms of solvent free energy with molecular dynamics (MD) simulations\cite{Matubayasi}(and references wherein). 
An additional caveat of this formulation was that the intramolecular packing term $E_{Intra}(S)$ and the sovlation free energy term $\Delta\mu$ were calculated independently, thus making the variance of the free energy calculation being the sum of these two parts. A formulation with unified calculation of both terms would therefore be highly desired, where firstly direct comparison with experimental data become more straight forward, and secondly variance increase due to independent calculations could be reduced.   

Machine learning, especially deep learning has been demonstrated to facilitate understanding of complex molecular systems\cite{Mart}. The caveat of brute force utilization of deep learning was that deep neural network worked as a black box of function composer in many cases and did not necessarily improve our fundamental physical understanding. Therefore, it was highly desirable to develop theoretical framework that was amenable to machine (deep) learning, and to achieve a good balance of explainability and prediction capacity. With present solvation free energy framework, it was rather difficult to utilize the optimization capacity of machine learning algorithms. 

We therefore proposed in this paper an alternative theoretical framework, a GSFE framework. The aim was to provide convenience for taking advantage of optimization capacity of machine learning algorithms on the one hand, and to support physical explanation (at least to some extent) on the other hand.      

\subsection{The generalized solvation free energy framework}
The definition of solute and solvent in molecular systems, while follow some intuition, was fundamentally arbitrary. The idea of the generalized solvation free energy (GSFE) framework was to define each basic physical comprising unit of a give complex system as solute and all its surrounding units as its specific solvent. The GSFE framework had the following basic properties: 

i) It was intrinsically multi-scale. Using protein molecular systems as examples, basic physical comprising units of which can be atoms, atomic groups, residues, clusters of residues, structural domains or even individual protein chains in mega protein complexes.

ii) Each basic comprising unit was both a solute and a comprising unit of solvent for its neighboring units simultaneously.

iii) Solvent was usually heterogeneous and specific for each solute unit.

\begin{figure}
\centering
  \includegraphics[width=10cm]{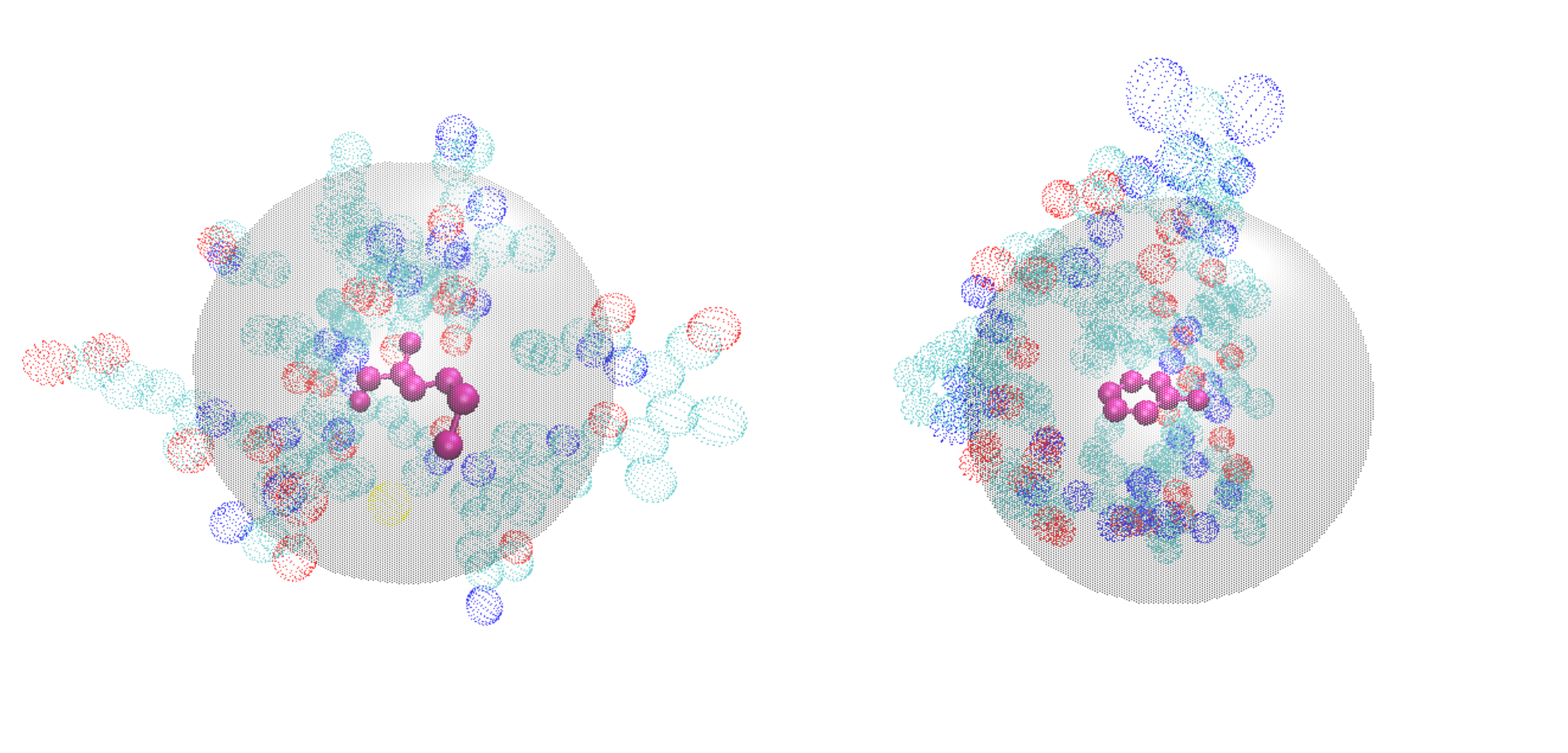}
  \caption{A schematic representation of the GSFE framework. Left: a residue (purple) was selected as a basic unit and its surrounding residues within certain cutoff distance were its specific solvent. Right: an atomic group (purple) was selected as a basic unit.}
  \label{fig:01}
\end{figure}
This was in stark contrast to traditional definition of solution where each basic unit/molecule was either solute or solvent, but usually not both. Solvent in traditional solution concept was in most cases homogeneous. A schematic illustration of the GSFE idea was presented in Figure \ref{fig:01}. The definition seemed to provide great difficulty in theoretical formulations as each solute unit has its own specific solvent, rendering representation of almost infinitely complex combinations of solvent impossible in multiple component complex molecular systems. Pondering further, this seemingly difficulty provided great convenience for machine learning of the free energy for the concerned molecular system as described below.

Again using a typical protein molecular system as an example with protein residues, water molecules and various ions as basic units. For a given residue, its specific solvent environment included all other surrounding residues, and water molecules and ions if it was on surface. However, it was quite well understood that in liquids and condensed soft matter, molecular interactions are usually quite short-ranged (treatment of long-ranged interactions were beyond the scope of this paper and would be investigated in the future). We therefore may safely restrict our attention to a limited spatial range by cut off the specific environment of a given solute. The task of properly describing solvent of a given solute unit remains daunting after this cutoff simplification. Specifically, two layers of neighboring residues for a fully buried solute residue could amount to 20 or even more. This would generate up to $20^{20}$ possible residue combinations with immense additional spatial variations. For solute residues that are close to surface, inclusion of complex possible combinations of water molecules and various other molecular entities (either explicitly or implicitly) were essential. Nevertheless, when putting aside the complexity of local solvent environment of a given solute for the time being, a great property emerged immediately for the GSFE framework. The probability of a given unit existed in a given solvent environment was solely determined by that environment. Therefore, all correlations among various comprising units were covered by the complexity of complex and heterogeneous solvent environment and was not necessary to be accounted for explicitly! The free energy of a given $n$-residue protein sequence $X=\{x_1,x_2,\cdots,x_n\}$ adopting a given structure (that defines solvent environment for each solute residue) may be written as:  
\begin{align}
F = - lnP(Structure|X)
\label{eqn:freeenergy}
\end{align}

By Bayes formula:
\begin{align}
P(Structure|X) &  = \frac{P(X|Structure)P(Structure)}{P(X)}\\ 
                        &  \propto P(X|Structure)P(Structure)
\label{eqn:bayes}                                  
\end{align}

When considering a maximum likelihood treatment, we focused our attention to the likelihood, which could be expanded as the following:
\begin{align}
P(X|Structure) \approx \prod_{i=1}^np(x_i|solvent_{x_i})
\label{eqn:likelihood}
\end{align}    

\begin{align}
lnP(X|Structure) \approx \sum_{i=1}^nlnp(x_i|solvent_{x_i})
\label{eqn:likelihood2}
\end{align} 
The approximation (besides maximum likelihood estimation) introduced in equations \ref{eqn:likelihood} and \ref{eqn:likelihood2}  were the interactions between the $i$th residue and all molecules not accounted for by $solvent_i$.

For the $i$th residue:
\begin{align}
p(x_i|solvent_i) = \frac{p(x_i, solvent_i)}{p(solvent_i)}
\label{eqn:conditionp}
\end{align}  
\begin{align}
p(solvent_i) = \sum_{AA_i = AA_1}^{AA_i = AA_{21}}p(AA_i|solvent_i)
\label{eqn:solventi}
\end{align}
Here $AA_i(i = 1,2,\cdots,21)$ represent 20 natural amino acids and all irregular residue was classified as the 21$st$ residue. The second major approximation was introduced here by forcing $p(solvent_i) = 1$ for any solvent environment belong to any basic unit $i$, the conditional probability and the joint probability term in equation \ref{eqn:conditionp} then became the same quantity. This approximation effectively assimilated the $P(Structure)$ term in \ref{eqn:bayes} into the corresponding likelihood term. All possible complex correlations between solute $i$ and its environment $solvent_i$ may be tackled by a neural network that was extremely good at solving non-linear mapping. Pairwise interaction approximation has been widely utilized in both KB and PB potentials. In GSFE framework, all correlations within selected solvent were included and relative importance of specific ordered interactions (pair, triple and higher ordered) were determined by the optimization process of the chosen machine learning algorithm.  

\subsection{A simple neural network implementation}
\subsubsection{Data sets}
We constructed our training set from the Cullpdb dataset \cite{wang2003pisces} that was generated on 2018.11.26. The sequence identity percentage cutoff was 25\%, the resolution cutoff was 2.0 angstroms, and the R-factor cutoff was 0.25. There were 9311 chains in the Cullpdb list. After downloaded the corresponding structures from the PDB (Prodein Data Bank, wwpdb.org), we culled the Cullpdb dataset and CASP13 dataset using CD-HIT server \cite{Ying2010CD}. After sequences that had more than 25\% identity in the two datasets were removed, we had 8129 structures from the Cullpdb dataset and 16 structures from the CASP13 dataset. We further devided the selected Cullpdb structures into a training set (7316 structures), a validation set (406 structures) and a test set (407 structures, referred to as the Cullpdb test set later) arbitrarily. 16 proteins of CASP13 and their decoy structures were utilized as one of the test set (with the top candidate structure from each competing group selected as decoy, so the number of decoys was the same as the number of participating groups for that target). This and four other data sets were utilized to test our model's ability in distinguishing native structures from decoys, the other four data sets were CASP5-CASP8 \cite{Rykunov2010New}, CASP10-13\cite{Yu2019}, I-TASSER \cite{zhang2010distance} and 3DRobot \cite{deng20153drobot} decoy sets. All of which were used to test ANDIS by Yu \cite{Yu2019}. The Rosetta dataset was also tested by Yu\cite{Yu2019}. However, we were not able to access it as the link was not available anymore.

\subsubsection{The input feature tensor}
We selected solvent residues of a given target residue according to its distance and backbone adjacency to the target residue as described below. Distances between two respective $C_\alpha$ atoms were utilized to represent residue distance. After calculating distances between all pairs of residues, the nearest 16 residues were save for each target residue as its solvent residues. These selected neighboring residues of a given target residue were divided into two categories, adjacent ($abs(ID(T)-ID(Neighbor)) <= 6$) and nonadjacent ($abs(ID(T)-ID(Neighbor)) > 6$) ones. $ID(T)$ was the sequence number of the target residue and $ID(Neighbor)$ was the sequence number of a neighboring residue, and $abs()$ function returned the absolute value. Adjacent residue were sorted with increasing $ID(T) - ID(Neighbor)$ order as in (-6, -5, -4, -3, -2, -1, 1, 2, 3, 4, 5, 6). Only adjacent residues belonged to the nearest 16 residues were included as neighbors. If a position in this list was not included as a neighboring residue, its features would be padded with zero. Nonadjacent residues were sorted in increasing order with their distances to the target residue, and stored in the nonadjacent list $(1, 2, 3, 4, 5, 6, 7, 8, 9)$, again empty positions were padded with zero. 

\begin{figure*}
\centering
  \includegraphics[height=4.5cm]{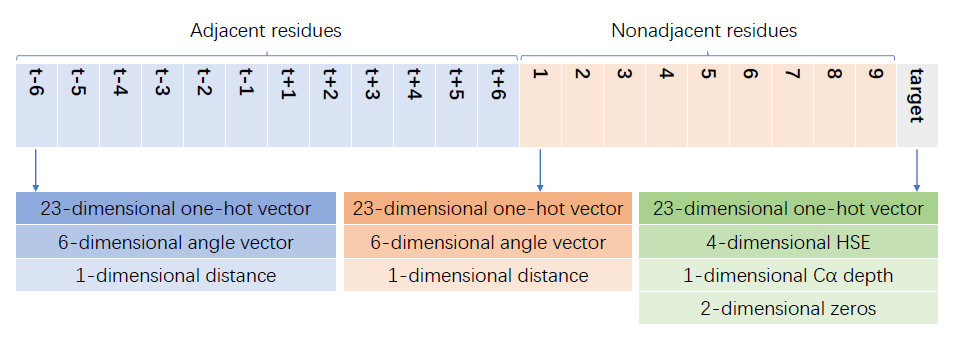}
  \caption{Feature inputs organization. Sequentially adjacent neighboring residues were indicated in blue, Sequentially nonadjacent ones were indicated in orange, and the information for target residue was indicated in green.}
  \label{fig:inputs}
\end{figure*}

After selecting specific solvent residues for each target residue, we defined the solvent of a target residue by identity, position and orientation of its selected neighboring residues. Each position in the both adjacent and nonadjacent list accommodated features of a solvent residue. Identity of each solvent residue was represented by a 23-dimensional one-hot vector (besides 20 regular residue types and 1 miscellaneous type that accounting for all non-regular residues, we added one type for non-neighboring residue and one type for target residue). Relative orientation for a pair of residues $A$ and $B$ were defined by six angles. We first calculated $C_\alpha(A)$-$C_\alpha(B)$, $C_\alpha(A)$-$N(A)$, $C_\alpha(A)$-$C_\beta(A)$ and $C_\alpha(A)$-$C(A)$ vectors respectively. Three angles were calculated for the following three vector pairs, $C_\alpha(A)$-$C_\alpha(B)$ and $C_\alpha(A)$-$N(A)$, $C_\alpha(A)$-$C_\alpha(B)$ and $C_\alpha(A)$-$C_\beta(A)$, and $C_\alpha(A)$-$C_\alpha(B)$ and $C_\alpha(A)$-$C(A)$. $N(A)$ and $C(A)$ were the nitrogen and carbon atoms of the peptide bond. Three additional angles were defined similarly with the order of $B$ and $A$ reversed. So for a pair of residues we utilized six angles to represent their relative orientation. Since GLY did not have $C_\beta$ atom, we utilized CHARMM\cite{CHARMM} parameter to derive its $HA2$ position based on its $C_\alpha$, $C$ and $N$ atoms. These six angle values followed the one hot identity vector. Finally, the distance to the target residue was added in as the last piece of information for a given solvent residue. (see Figure.\ref{fig:inputs}). Therefore, each neighboring solvent residue was represented by 30 values (23 for one-hot identity encoding, 6 for orientation and 1 for distance) in the input vector.

Target residue in the input vector was represented by a 23-dimensional one-hot vector, residue depth (represented by distance of its $C_\alpha$ atom from solvent accessible protein surface) and Half Sphere Exposure (HSE) of the target residue were also included in input features. HSE was a two-dimensional measure of solvent exposure. Essentially, the number of $C_\alpha$ atoms around a residue in the direction of its side chain and in the opposite direction (within a radius of $13 \AA$) was counted. HSE came in two flavors: $HSE_\alpha$ and $HSE_\beta$. The former only utilized $C_\alpha$ atom positions, while the latter utilized both $C_\alpha$ and $C_\beta$ positions. To make its number of values even with neighboring solvent residues, two zeros were padded. (see Figure \ref{fig:inputs}). 

In summary, for each target residue we had vector encodings for 21 neighboring residues (12 adjacent and 9 nonadjacent) and the target residue, they form a 660-dimensional feature vector (see Figure \ref{fig:inputs}), which served as the input of a neural network. All features were calculated with the Biopython library. Input vector for all target residues of a protein chain was fed to the neural network (described below) as the input tensor.

\subsubsection{Neural network architecture and training process}
\begin{figure}
\centering
  \includegraphics[height=3.5cm]{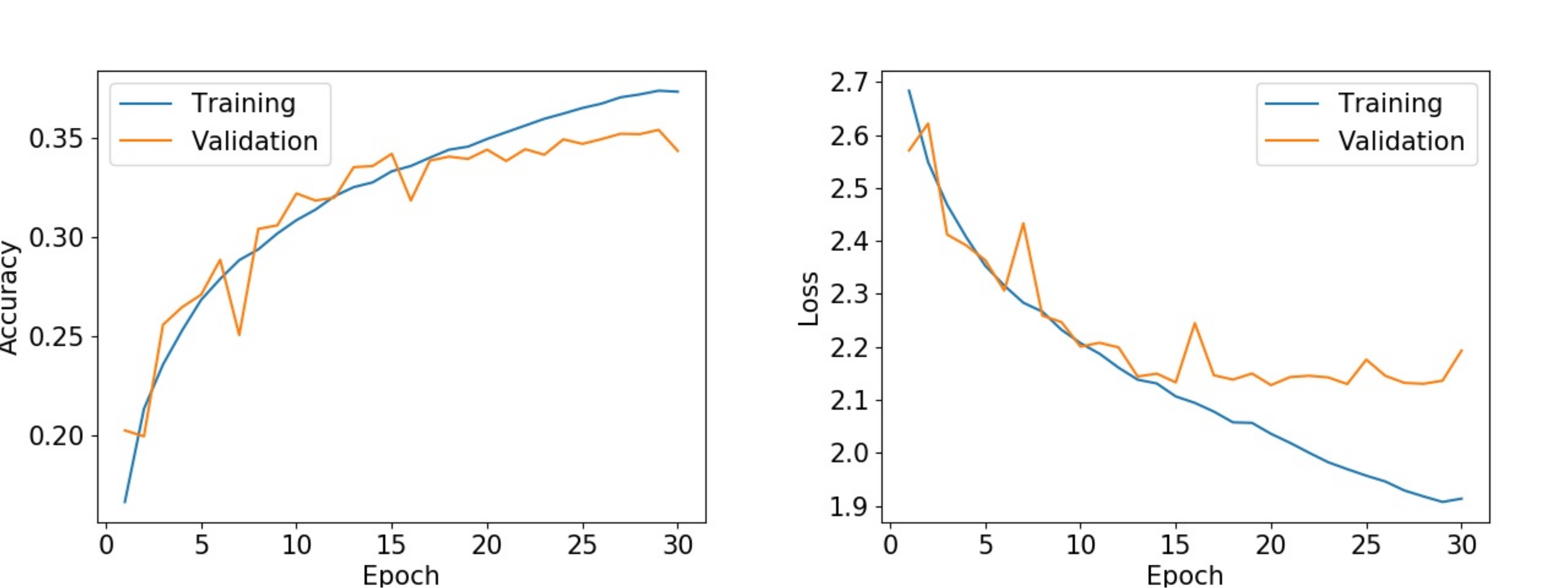}
  \caption{Accuracy (left) and Loss value (right) of the training dataset and validation dataset.}
  \label{fgr:fig2}
\end{figure}

A simple four layer feedforward neural network was constructed with PyTorch. There were three hidden layers, each with 512 neurons. ReLU was the selected activation function for all layers except the output layer, for which softmax activation was utilized. The network had 21 neurons in the output layer, 20 regular residues were classified as the first 20 classes and all other non-standard residues were classified as the 21-$st$ class. The cross-entropy was utilized as the loss function. For each target residue the neural network took its 660-dimensional input vector and predicted its residue type. The network was trained with our training dataset. The Stochastic Gradient Descent (SGD) optimizer was selected with a learning rate of 0.1. For each batch we trained all target residues in a protein chain. The network was trained for 30 epoches and was tested with the validation set after each epoch. Loss value and accuracy of each epoch were shown in Fig \ref{fgr:fig2}. Best performing parameters in the validation set were saved as our final model. Finally we tested the model with the Cullpdb test set and obtained a target identity prediction accuracy of 35.2\%. The training was carried out on a GeForce RTX 2080Ti GPU and costed approximately 8 minutes for one epoch.

\section{RESULTS}
To evaluate a candidate structure, we calculated the log likelihood of each residue with our trained neural network and obtained the log likelihood of the whole structure simply by summation of log likelihood of each residue according to equation \ref{eqn:likelihood2}. Larger likelihood value corresponding to lower free energy or more native-like structure. We tested our model's ability in selecting native structure from decoys on the four datasets described in the method section. These datasets have been used by Yu to compare ANDIS with eight other methods \cite{Yu2019}, most of which were atomic KB potentials. The results from Yu's work and our evaluation results were listed in table \ref{tb:tb1}. The comparison suggested that our simple residue position and orientation based GSFE neural network implementation was competitive in distinguishing native structures from decoys. We further tested our model on 16 protein of the constructed CASP13 decoy set, in which all sequences had 25\% or lower sequence identity with any sequences in our training set. As shown in table\ref{tb:tb2}, we recognized 8 out of 16 target structures from decoys and the average Z-score was 1.27. The rank of native structures for unsuccessful targets were quite high as well.

\begin{table}[h]
\small
  \caption{\ Performance comparison in native state recognition. The number of proteins whose native structure was given the lowest free energy score (our method use the largest likelihood) by respective potential were listed outside the parentheses. The average Z-scores of native structures were listed in parentheses. Z-score was defined as $(<E_{decoy}> - E_{native} )/\delta $ (our method $(E_{native}-<E_{decoy}>)/\delta$), where $E_{native}$ is the energy score of native structure, $<E_{decoy}>$ and $\delta$ were the average and the standard deviation of energy scores respectively for all decoys in the set.}
  \label{tb:tb1}
  \begin{tabular*}{0.48\textwidth}{@{\extracolsep{\fill}}lllll}
    \hline
    Decoy sets     & CASP5-8  & CASP10-13   & I-TASSER  & 3DRobot \\
    \hline
    No. of targets & 143 (2759) & 175 (13474) & 56 (24707) & 200 (60200) \\
    Dfire          & 64 (0.61)  & 56 (0.72)   & 43 (2.80)  & 1 (0.83) \\
    RW             & 65 (1.01)  & 36 (0.86)   & \textbf{53} (4.42)  & 0 (-0.30)\\
    GOAP           & 106 (1.67) & 89 (1.62)   & 45 (4.98)  & 94 (1.85)\\
    DOOP           & 135 (1.96) & 121 (1.99)  & 52 (6.18)  & 197 (3.53)\\
    ITDA           & 71 (1.15)  & 117 (1.67)  & 52 (4.98)  & 196(3.83) \\
    VoroMQA        & 132 (2.00) & 111 (1.77)  & 48 (5.11)  & 114 (1.89) \\
    SBROD          & 88 (1.62)  & 119 (\textbf{2.32})  & 33 (3.25)  & 49 (1.76) \\
    AngularQA      & 59 (1.26)  & 24 (1.11)   & 29 (1.82)  & 9 (0.99) \\
    ANDIS          & 138 (\textbf{2.16}) & 129 (\textbf{2.32})  & 47 (\textbf{6.45})  & \textbf{200} (\textbf{4.99}) \\
    GSFE           & \textbf{139} (2.01) & \textbf{136} (1.87)  & 46 (4.13)  & \textbf{200} (3.50) \\
    \hline
  \end{tabular*}
\end{table}

\begin{table}[h]
\small
  \caption{\ CASP13 result, native structure rank in decoy structures.}
  \label{tb:tb2}
  \begin{tabular*}{0.48\textwidth}{@{\extracolsep{\fill}}llll}
    \hline
    Structure id   & Rank (Z-score) & Structure id  & Rank (Z-score) \\
    \hline
    T0950-D1    & 1/39 (2.65)  & T0966-D1    & 1/87 (1.18) \\
    T0953s1-D1  & 4/90 (1.92)  & T0968s1-D1  & 1/94 (1.42) \\
    T0954-D1    & 1/87 (1.33)  & T0968s2-D1  & 2/95 (1.07) \\
    T0955-D1    & 18/92 (0.68) & T1003-D1    & 8/89 (1.13) \\
    T0957s1-D1  & 2/92 (1.01)  & T1005-D1    & 1/83 (1.23) \\
    T0957s2-D1  & 1/91 (1.60)  & T1008-D1    & 25/91 (0.84) \\
    T0958-D1    & 4/90 (1.22)  & T1009-D1    & 1/85 (1.04) \\
    T0960-D1    & 22/84 (0.71) & T1011-D1    & 1/82 (1.33) \\
    \hline
  \end{tabular*}
\end{table}

\section{DISCUSSION}
It was important to note that the GSFE formulation had its own limitation. It was trained on available structures and therefore was likely to be not as effective under physical conditions that were far from those generating these data. Traditional formulation was more powerful in many interesting schemes. For example, if we were interested in investigating the response of a protein under periodically varying electric fields; or if we were interested in interfacial properties. 

In this simple neural network implementation of GSFE framework, we utilized residue level spatial resolution with orientation of side chain represented by $C_\beta$ (or $HA2$ for GLY). Despite the simplicity, the trained model were competitive in selecting native structures from decoys when compared with sophisticated atomic potentials. The weakness of the trained model as reflected by relatively small value of Z scores was likely due to the fact that only native structures were utilized for the training, and this issue would be tackled in our future work. 

The free energy resolution and spatial range of molecular interactions of GSFE was apparently limited by the specific basic unit selected and available data. If we chose to carry out atomistic GSFE training with a large cutoff distance, the complexity of such local solvent would require much larger data sets and longer training time. However, due to the intrinsic multi-scale definition, it was hopeful that we might train hierarchical GSFE models to account for both relatively long-range interactions and local chemical details. This was also a future direction of our GSFE development.

At first sight, the possible local chemical environment for any residue/atom (and other possible definition of basic unit) was an immense space and the available data sets (structures in protein data bank) seemed hopelessly small. However, evolution in billions of years by huge number of organisms had sampled and found at least a significant fraction, if not all, of important local chemical environment spaces that were partially represented in present available data sets. Nevertheless, the approximation of forcing $p(solvent_i) = 1$ had room to be improved. The $P(Structure)$ term in equation\ref{eqn:bayes} was left out in the maximum likelihood treatment and was effectively assimilated into the local likelihood ($p(x_i|solvent_i)$) by the above approximation. Physically, the $P(Structure)$ could be expressed as the following:
\begin{align}
P(Structure) = P(Structure|Fold)P(Fold)
\end{align}
With $P(Fold)$ being the probability of concerned protein fold in the manifold of all possible protein folds. Apparently, accurate quantization of this equation necessitated understanding of the whole protein fold space on the one hand, and structural variation within a specific fold on the other hand. It was necessary to address this issue properly so that we may move from present maximum likelihood estimation to full Bayesian treatment. This was actually a long term goal of GSFE development.

\section{Conclusions}
We proposed the GSFE framework to achieve a balance of physical interpretability and powerful non-linear optimization capacity of machine learning. This framework was intrinsically multi-scale and amenable to various machine learning optimization. One distinctive feature of GSFE was that local correlations accounted for were not limited to pairwise interactions, which was a widely utilized practice in many KB and PB potentials. A simple neural network implementation of GSFE at residue level for protein structural model assessment was found to be competitive with sophisticated state-of-the-art atomic KB potentials in distinguishing native structures from decoys. In future work, we planned to carry out investigations on more sophisticated neural network architectures and training strategies to achieve strong capability in ranking decoys besides distinguishing native structures from decoys, and to develop hierarchical implementations of GSFE.
  
\section*{Conflicts of interest}
There were no conflicts to declare.

\section*{Acknowledgements}
This work has been supported by the postdoctoral start up fund from Jilin University (801171020439), by National Natural Science Foundation of China (31270758), and by the Fundamental Research Funds for the Central Universities (451170301615).



\balance


\bibliography{./gsf1} 
\bibliographystyle{rsc} 

\end{document}